\documentstyle[11pt,epsf,rotate,aaspp4]{article}

\begin{document}

\newcommand\bb[1] {   \mbox{\boldmath{$#1$}}  }

\newcommand\del{\bb{\nabla}}
\newcommand\bcdot{\bb{\cdot}}
\newcommand\btimes{\bb{\times}}
\newcommand\vv{\bb{v}}
\newcommand\B{\bb{B}}
\newcommand\BV{Brunt-V\"ais\"al\"a\ }
\newcommand\iw{ i \omega }
\newcommand\kva{ {k v_A}  }
\newcommand\kb{ \bb{k\cdot b}  }
\newcommand\kkz { \left( \frac{k}{k_Z}\right)^2\>}

    \def\dd{\partial}
    \def\tilde{\widetilde}
    \def\etal{et al.}
    \def\eg{e.g. }
    \def\etc{{\it etc.}}
    \def\ie{i.e.}
    \def\beq{ \begin{equation} }
    \def\eeq{ \end{equation} }
    \def\spose#1{\hbox to 0pt{#1\hss}} 
    \def\ltsim{\mathrel{\spose{\lower.5ex\hbox{$\mathchar"218$}}
	 \raise.4ex\hbox{$\mathchar"13C$}}}

\def\tilde{\widetilde}

\newcommand{\schwz}{ {\dd  \ln P\rho ^{-\gamma} \over \dd Z}}
\newcommand{\schwR} { {\dd  \ln P\rho ^{-\gamma} \over \dd R} }

\newcommand{\balbz}{ {\dd  \ln T \over \dd Z}}
\newcommand{\balbR} { {\dd  \ln T \over \dd R} }

\long\def\Ignore#1{\relax}

\title{Black Hole Accretion Disks On The Edge}

\author{Kristen Menou\altaffilmark{1} }
\affil{Virginia Institute of Theoretical Astronomy, Department of Astronomy,}
\affil{University of Virginia, P.O. Box 3818, Charlottesville, VA 22903, USA}
\altaffiltext{1}{Celerity Foundation Fellow}

\begin{abstract}

The local axisymmetric stability of hydrodynamical and magnetized,
nearly-Keplerian gaseous accretion disks around non-rotating black
holes is examined in the vicinity of the classical marginally-stable
orbit (at radii $ \sim R_{\rm ms}$). An approximate Paczynski-Wiita
pseudo-Newtonian potential is used. Hydrodynamical disks are linearly
unstable inside a radius which differs slightly from the classical
$R_{\rm ms}$ value because of finite pressure and radial
stratification effects. Linear stresses associated with unstable
hydrodynamical modes vanish exactly at the radius of marginal
stability and are generally positive inside of that radius. When a
magnetic field is introduced, however, the concept of radius of
marginal stability becomes largely irrelevant because there are
linearly unstable magneto-rotational modes everywhere. Associated
linear stresses are positive and continuous across the region of
hydrodynamical marginal stability, even for large-scale "hydro-like"
modes subject only to weak magnetic tension. This conclusion is valid
for arbitrarily thin disks (in ideal MHD) and it does not require a
large-scale "radially-connecting" magnetic field.  Results on
hydrodynamical diskoseismic modes trapped in deep relativistic
potential wells should be revised to account for the short-wavelength
Alfv\`en-like behavior of inertio-gravity waves in magnetized disks.

\end{abstract}


\section{Introduction}

Over the past few years, the question of the nature of accretion in
the vicinity of black holes has received considerably renewed
attention. This interest stems from a combination of new observational
data and a deeper theoretical understanding of the accretion
problem. On the observational side, significant progress has been made
since a new generation of powerful X-ray telescopes (Chandra and
XMM-Newton) became available. These instruments have been used to
strengthen earlier claims of evidence for general relativistic effects
arising deep in the potential well of spinning black holes, in the
form of broad iron emission line profiles (Iwasawa et al. 1996;
Dabrowsky et al. 1997; Branduardi-Raymont et al. 2001; see also
Reynolds \& Begelman 1997; Young, Ross \& Fabian 1998). In addition,
it has been suggested that in some systems one actually witnesses
black hole spin energy being tapped to power part of the observed
X-ray emission (Wilms et al. 2001, Miller et al. 2002).

On the theoretical side, efforts to reconsider classical results (see,
e.g., Novikov \& Thorne 1973; Page \& Thorne 1974, Shapiro \&
Teukolsky 1983) on the marginally-stable orbit and the zero-torque
boundary condition at $R_{\rm ms}$ have been motivated by the
compelling case made for magnetic fields and MHD turbulence being the
driving mechanism behind accretion (Balbus \& Haley 1991; 1998; Balbus
2003; Blaes 2003). Recent theoretical efforts on near-black-hole
accretion have focused on analytical models of steady-state accretion
(Krolik 1999; Gammie 1999; Agol \& Krolik 2000; Paczynski 2000;
Afshordi \& Paczynski 2003) or on global numerical simulations of
magnetized flows in an approximate pseudo-Newtonian potential for
non-rotating black holes (Hawley \& Krolik 2001; 2002; Reynolds \&
Armitage 2001, Armitage, Reynolds \& Chiang 2001; Krolik \& Hawley
2002).

Gammie \& Popham (1998; see their footnote 7) and Araya-Gochez (2002)
have studied some important aspects of the linear stability problem
for magnetized disks around black holes in full general relativity,
but it seems that there has been no attempt to use this approach to
address some of the debated points in the literature on the subject of
near-black-hole accretion (e.g. validity of zero-torque condition,
role of disk thickness, differences between hydrodynamical and
magnetized disks). This is undoubtedly because there are issues that a
local, linear analysis cannot fully address and which require global
numerical simulations. Nonetheless, as I argue below, such an analysis
still provides valuable insights on some of the key aspects of the
accretion problem in the vicinity of a black hole. In \S2, I describe
the formalism used for the linear analysis. I present a comparison of
results for hydrodynamical and magnetized thin accretion disks in
\S3. Implications and limitations of this work are discussed in \S4.

\section{Formalism}

We work in cylindrical coordinates $({R}, {\phi}, {Z})$. Without loss
of generality, we restrict our analysis to a basic state disk with a
purely vertical magnetic field, $\bb{B}=(0,0,B)$, and a rotation law,
$\bb{\Omega} = (0, \Omega(R),0)$, that is constant on cylinders.  We
consider axisymmetric Eulerian perturbations (denoted by a prefix
$\delta$) with WKB space-time dependence $e^{i({k Z} - \omega t)}$.
We work in ideal MHD (i.e. microscopic diffusion processes are
ignored) and we neglect weak circulations of the basic state, such as
that caused by a finite fluid viscosity.

The formalism used for this analysis is essentially the same as the
one developed by Balbus \& Hawley (1991; 2002). Standard MHD equations
and expressions for the leading-order WKB terms (within the Boussinesq
approximation) can be found in the first of these two references. Two
of these equations for linear perturbations are useful for the
discussion below and they are reproduced here:

\beq  \label{eq:one}
-\iw  \delta v_\phi + \delta v_R\, {\kappa^2 \over 2 \Omega}
-i {k B} {\delta B_\phi \over 4 \pi \rho} = 0,
\eeq

\beq \label{eq:two}
-\iw  \delta B_R - i {k B} \delta v_R = 0.
\eeq

Linear perturbations satisfy the dispersion relation

\begin{eqnarray} 
{\tilde\omega}^4 -
{\tilde\omega}^2 (N^2+ \kappa^2)
- 4 \Omega^2 (\kva)^2 =0, \label{eq:ds}
\end{eqnarray}

where

\begin{eqnarray}
{v_A} = { {B}/\sqrt{4\pi\rho}}, \qquad {\tilde\omega}^2=\omega^2-(\kva)^2,
\end{eqnarray}

and $\rho$ is the mass density of the fluid. Unstable modes are best
characterized by their growth rate, $\gamma =-i \omega$.  The \BV
frequency and the epicyclic frequency appearing in the dispersion
relation are defined, respectively, by

\beq
N^2 =  -{3 \over 5 \rho} {d  P \over d R} 
{d  \ln P\rho ^{-5/3} \over d R}, \qquad \kappa^2 = 
{1\over R^3} {d R^4\Omega^2\over d R},
\eeq

where $P$ is the gas pressure and $5/3$ is the adiabatic index of the
gas.

Following Balbus \& Hawley (2002), the growth rate, $\gamma$, of
unstable modes obeys

\beq \label{eq:growth}
\gamma^2= -(\kva)^2 - \frac{1}{2} \left[  N^2 + 
\kappa^2 -\sqrt{(N^2+\kappa^2)^2+16 \Omega^2 (\kva)^2}  \right].
\eeq

The wavenumber ($k$) and growth rate of the fastest growing mode are
given by

\beq
(\kva)^2_{\rm max}= \Omega^2 \left( 1-\frac{(N^2+\kappa^2)^2}{16 \Omega^4} \right) , \qquad \gamma_{\rm max}=\frac{\Omega}{4} \left( \left| \frac{d \ln \Omega^2}{d \ln R} \right| - 
\frac{N^2}{\Omega^2}  \right).
\eeq

The $R \phi$ component of the linear perturbation stress tensor, which
measures the angular momentum flux associated with unstable modes
(Balbus \& Hawley 2002; 1998), is

\beq
T_{R \phi}=\rho \left[  \delta v_R \delta v_\phi -\delta v_{AR} \delta v_{A \phi} \right],
\eeq
where the first term is the Reynolds stress and the second term is the Maxwell stress:

\beq
\rho \delta v_R \delta v_\phi = \rho (\delta v_R)^2 \frac{\Omega}{D \gamma} 
\left[ \frac{(\kva)^2}{\gamma^2} \left| \frac{d \ln \Omega}{d \ln R}\right| - \frac{\kappa^2}
{2 \Omega^2}\right] ,
\eeq

\beq
-\rho \delta v_{AR} \delta v_{A\phi} = \rho (\delta v_R)^2 \frac{2\Omega}{D \gamma} 
\frac{(\kva)^2}{\gamma^2}, 
\eeq

and $D=1+(\kva / \gamma)^2$. In what follows, we will focus on the
dimensionless stress

\beq
T^*_{R\phi} = \frac{T_{R \phi}}{\rho (\delta v_R)^2}.
\eeq

While Balbus \& Hawley (2002; see also Narayan et al. 2002) were
interested in the general stability and transport properties of
accretion flows with strongly destabilizing radial stratifications
($N^2 \sim - \Omega^2$), our specific interest here is in the possible
change of stability properties caused by the gravitational potential
of a non-rotating black hole in the vicinity of the classical
marginally stable orbit ($R_{\rm ms}$).  We approximate the
Schwarzschild metric with the pseudo-Newtonian potential of Paczynski
\& Wiita (1980), which implies a Keplerian rotation law

\beq
\Omega^2_K(R) =\frac{GM_{\rm BH}}{R(R-R_g)^2},
\eeq

where $G$ is the gravitational constant, $M_{\rm BH}$ is the central
black hole mass and $R_g=2 GM_{\rm BH}/c^2$ is the Schwarzschild radius
of the non-rotating hole ($c$ is the speed of light). This approximate
potential is known to reproduce accurately the location of the
classical marginally-stable ($R_{\rm ms}=3 R_g$) and marginally-bound
($R_{\rm mb}=2 R_g$) orbits.  We focus our attention on the stability
of nearly-Keplerian gaseous accretion disks, i.e. gaseous disks in
near rotational equilibrium.  In the absence of any accretion
(i.e. infall), the equilibrium (basic state) rotation law of such a
disk is governed by the radial component of the Euler equation,

\beq \label{eq:omega}
R \Omega^2 =R \Omega_K^2 +\frac{1}{\rho} \frac{d P}{d R}.
\eeq

The assumption of near Keplerianity means that radial pressure
gradients are small compared to the two other terms in the above
equation: $\Omega \simeq \Omega_K$. This is equivalent to requiring
that the disk geometrical thickness be small compared to its radius
($H/R \ll 1$). A further consequence of this assumption is that,
absent any variations of the density or temperature over a lengthscale
comparable to the disk pressure scale height, $H$, the epicyclic
frequency, $\kappa^2$, is close to the Keplerian value as well. In
what follows, we will neglect any such deviation of $\kappa^2$ from
local Keplerianity.

Since our basic state disk configuration is, by assumption,
non-accreting, it would seem natural to assume that any radial
stratification (that would in general arise from accretion and
associated dissipation) is absent. It is possible, however, to imagine
situations in which a non-accreting gaseous disk still possesses a
non-zero radial stratification, as could be the case for instance in
the presence of external sources of irradiation. For this reason, we
generalize our discussion to the cases with: (i) no radial
stratification ($N^2=0$), (ii) a stabilizing radial stratification
($N^2=0.1 \Omega^2$) and (iii) a destabilizing radial stratification
($N^2 =-0.1 \Omega^2$). The values $\pm 0.1 \Omega^2$ are
intentionally chosen to be large (for a thin disk), to better
illustrate the effects of finite $N^2$ values on stability.  {
Reducing these values to more realistic levels would not affect the
results qualitatively, but it would make the identification of the
effects on figures more difficult.}

\section{Results}

\subsection{Stability of Hydrodynamical Disks}

While the formalism described in the previous section has been
developed for magnetized disks, the hydrodynamical limit can be
recovered by enforcing $\kva = 0$. The dispersion relation reduces to
second order and the growth rate of unstable modes becomes independent
of wavenumber ($k$): $\gamma^2= -(N^2+\kappa^2) $.  One recognizes one
of the classical Solberg-H\o iland criteria for axisymmetric stability
(e.g. Tassoul 1978) emerging from this expression for
$\gamma^2$. Around a black hole, the Keplerian rotation law becomes
unstable inside of the classical marginally-stable orbit, $R_{\rm
ms}=3 R_g$, where $\kappa^2 < 0$ (for all wavenumbers $k$). While this
is an exact result for a test particle (or a dust disk),
hydrodynamical disks behave somewhat differently. First, the value of
the disk epicyclic frequency, $\kappa^2$, is slightly modified by the
presence a finite radial pressure gradient (eq.~[\ref{eq:omega}]), a
small effect that we have chosen to ignore for simplicity. Second, the
presence of radial stratification modifies the radius of marginal
stability (i.e. the radius at which $\gamma^2=0$). Marginal stability
occurs at a radius somewhat larger (smaller) than the standard $R_{\rm
ms}=3 R_g$ in the presence of a destabilizing (stabilizing) radial
stratification. These modifications remain small for cold/thin disks
(with $N^2 \ll \Omega^2$).

Dimensionless Reynolds stresses associated with linearly unstable
hydrodynamical modes take the (wavenumber independent) form: $T^*_{R
\phi}=-(\kappa^2/2 \Omega^2) \times (\Omega/\gamma)$. In the absence
of radial stratification, $\gamma=\sqrt{- \kappa^2}$, the
dimensionless stress is everywhere positive within the radius of
marginal stability and it vanishes exactly at that radius. When a
finite radial stratification is present, however, the degeneracy
between the growth rate, $\gamma$, and the epicyclic frequency,
$\kappa^2$, is broken, and $T^*_{R \phi}$ diverges at marginal
stability ($\gamma=0$ but $\kappa^2 \neq 0$). The divergence is
positive for a stabilizing radial stratification and negative for a
destabilizing one. Therefore, in a disk with $N^2 < 0$, there is a
small region with negative stresses right inside of the radius of
marginal stability. These results are graphically illustrated in
figure~\ref{fig:three} (see below).

The divergence of the dimensionless stress does not apply to the
physically-relevant dimensional stress,
however. Equation~(\ref{eq:one}) for linear perturbations shows that,
in a hydrodynamical disk (with $B=0$), when $\gamma=-i \omega
\rightarrow 0$ while $\kappa^2 \neq 0$, $\delta v_R$ must $\rightarrow
0$ as well. [This requirement disappears if $\kappa^2=0$ at marginal
stability (i.e. absent any radial stratification), but there is no
divergence of $T^*_{R \phi}$ in that case because $\delta v_R$ can
remain finite (see also Fig.~\ref{fig:three}).] The fact that $\delta
v_R \rightarrow \gamma \rightarrow 0$ at marginal stability is
precisely what makes the dimensionless stress, $T^*_{R \phi}$, diverge
(if $N^2 \neq 0$). On the other hand, the dimensional stress, $T_{R
\phi} \propto T^*_{R \phi} \times (\delta v_R)^2$, vanishes exactly at
marginal stability, where $\delta v_R \rightarrow \gamma \rightarrow
0$. All this guarantees that the physically relevant linear stress
always vanishes at marginal stability, whether there is radial
stratification or not.  Note that a similar situation occurs at
marginal stability ($\delta v_R \rightarrow \gamma \rightarrow 0$) in
a magnetized disk, as equation~(\ref{eq:two}) for linear perturbations
shows. Despite the possibility of divergence at marginal stability, we
found that using $T^*_{R \phi}$ remained very useful because of its
dimensionless nature.\footnote{Narayan et al. (2002) use the
dimensionless quantity $t_{r \phi}=T^*_{R \phi} \times (\gamma /
\Omega)$, which remains finite at marginal stability where the
dimensional stress, $T_{R \phi}$, vanishes (and $T^*_{R \phi}$
diverges). }

\subsection{Stability of Magnetized Disks}

\begin{figure}
\plotone{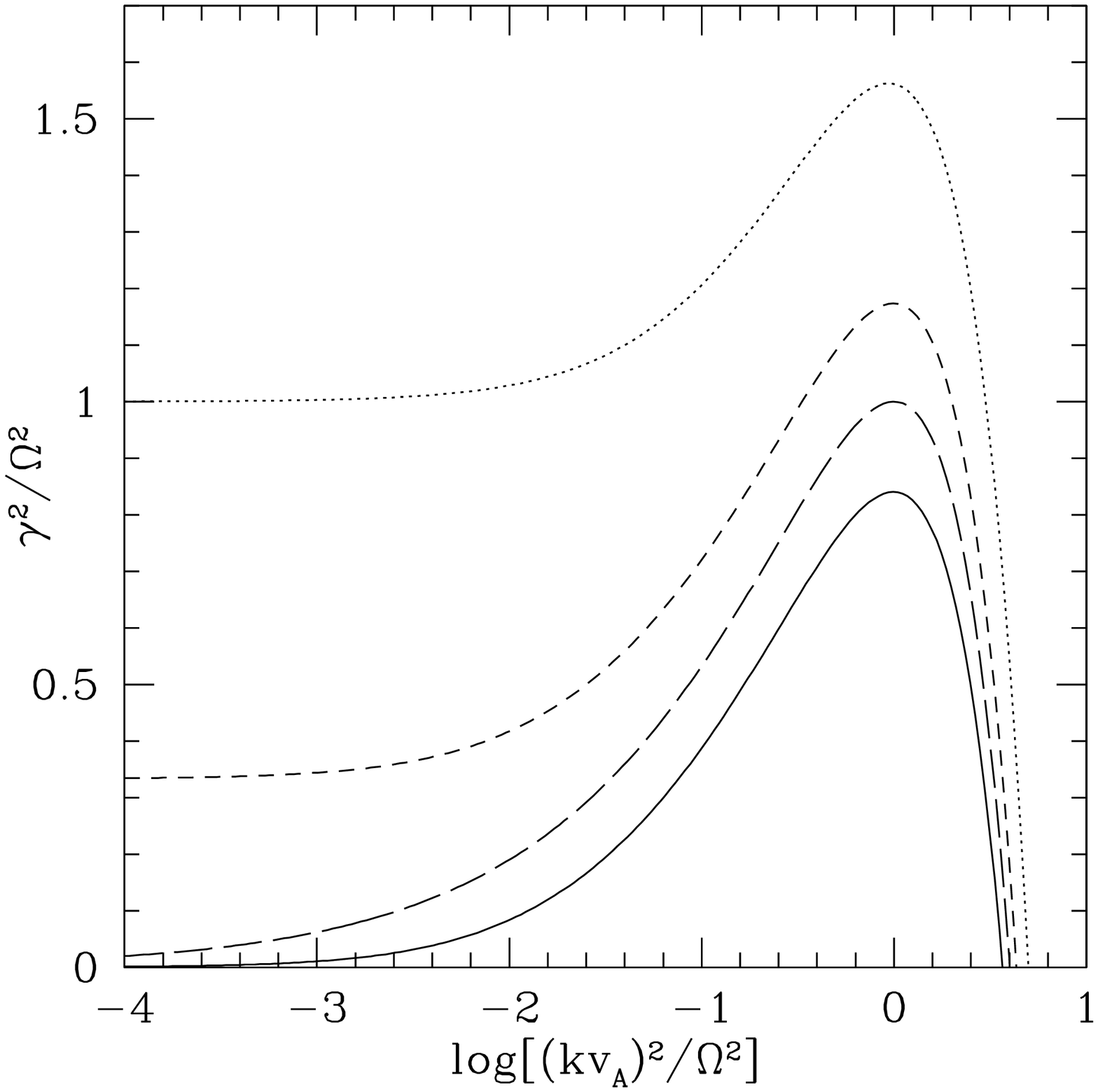}
\caption{Dimensionless growth rate, $\gamma / \Omega$, of linearly
unstable modes as a function of dimensionless wavenumber, $\kva /
\Omega$, in a magnetized disk without radial stratification. The
various curves correspond to different locations in the vicinity of
the classical marginally-stable orbit at $R_{\rm ms}=3 R_g$ (solid:
$4 R_g$; long-dashed: $3 R_g$; short-dashed: $2.5 R_g$; dotted: $2
R_g$).
\label{fig:one}}
\end{figure}

Figure~\ref{fig:one} shows growth rates of unstable modes, as a
function of dimensionless wavenumber, $\kva / \Omega$, at four
different locations in the vicinity of the classical marginally stable
orbit, $R_{\rm ms}=3 R_g$. Only the case without radial stratification
($N^2=0$) is shown because it was found that a value of $N^2 <<
\Omega^2$ only slightly reduces or increases growth rates in an
approximately wavenumber independent way.  For each of the curves in
figure~\ref{fig:one}, the mode with maximal growth is easily
identified at $\kva \sim \Omega$. The growth rate of this most
unstable mode increases as one approaches the black hole horizon
because the shear and the epicyclic frequency become ever more
destabilizing. Values of the growth rate for the most unstable mode
shown in figure~\ref{fig:one} are found to be in very good agreement
with corresponding fully general relativistic calculations for
Schwarzschild black holes (Araya-Gochez 2002).

One observes a qualitative change in the behavior of small wavenumber
(large scale) modes as a function of radius. At radii $R \ge 3 R_g$
(for $N^2=0$ and a strictly Keplerian epicyclic frequency), the growth
rate of these modes tends to zero in the limit $\kva \rightarrow
0$. Within $3 R_g$, however, the growth rate tends to a finite value,
which is $\gamma=\sqrt{-\kappa^2}$. This is exactly the growth rate of
unstable modes in a hydrodynamical disk at the same location within
the radius of marginal stability. Therefore, large scale modes
characterized by weak magnetic tension (small $\kva$) behave
asymptotically like their hydrodynamical counterparts. Note that this
is the only physical significance attached to the hydrodynamical
radius of marginal stability in a magnetized disk. Because there
always are unstable modes, on scales small enough for magnetic tension
to be important, there is no effective radius of marginal stability in
a magnetized disk. Finite radial stratification or pressure effects
can modify somewhat the exact radius at which small wavenumber modes
become unstable (from the canonical $3 R_g$ value; see discussion of
Fig.~\ref{fig:three} below), but they do not qualitatively alter the
stability or stress properties.

\begin{figure}
\plotone{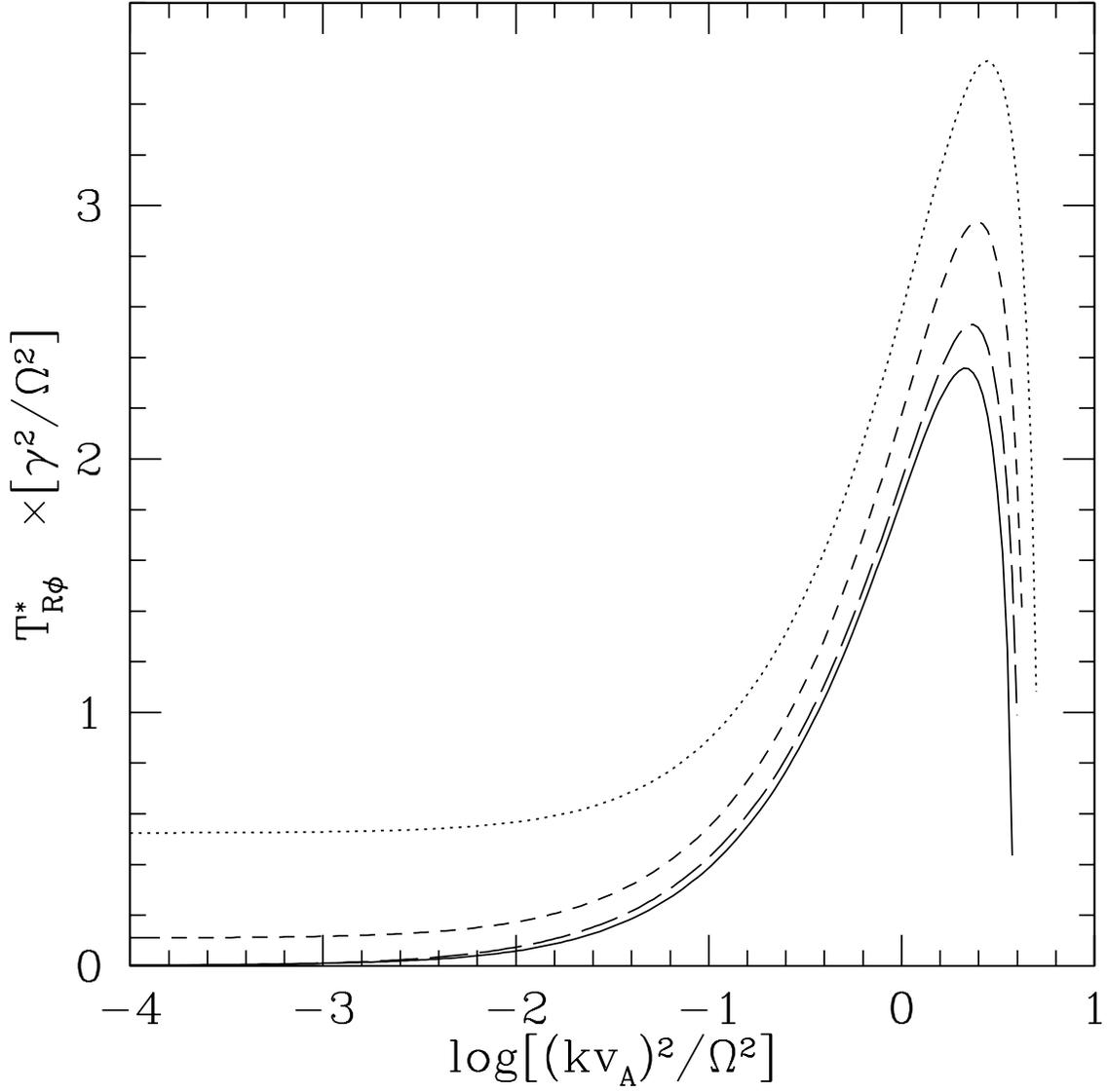}
\caption{Dimensionless stress, $T^*_{R\phi}$ ($\times
\gamma^2/\Omega^2$), of linearly unstable modes as a function of
dimensionless wavenumber, $\kva/\Omega$, in a magnetized disk without
radial stratification. Notation is identical to Fig.~1 (which shows
the corresponding growth rates).
\label{fig:two}}
\end{figure}

Figure~\ref{fig:two} shows the total (Reynolds + Maxwell)
dimensionless stresses, $T^*_{R \phi}$, associated with linearly
unstable modes, as a function of the dimensionless wavenumber, $\kva /
\Omega$, at the same four locations in the vicinity of the classical
marginally stable orbit (same notation and $N^2=0$ assumption as in
Fig.~\ref{fig:one}).  To illustrate the vanishing of the dimensional
stress at marginal stability, we have multiplied the dimensionless
stress by $\gamma^2/\Omega^2$ (in this figure only), since we know
that $\delta v_R \rightarrow \gamma \rightarrow 0$ at marginal
stability (see \S3.1).  Again, a qualitative change in the stress
across the hydrodynamical radius of marginal stability occurs only for
large scale modes. The stress is asymptotically zero outside of $3
R_g$ and asymptotically finite inside of $3 R_g$, in the limit $\kva
\rightarrow 0$.

Figure~\ref{fig:three} summarizes the picture that emerges for linear
perturbation stresses in magnetized and hydrodynamical disks in the
region around the classical marginally stable orbit. The stress for
the fastest growing mode in a magnetized disk is positive and
continuous throughout the region of interest. The introduction of a
stabilizing or destabilizing radial stratification does not
qualitatively change this result. On the contrary, the (wavenumber
independent) linear perturbation stress in a hydrodynamical disk is
positive only within the radius of marginal stability, whose location
is affected by the introduction of radial stratification (see lower
short- and long-dashed lines in Fig.~\ref{fig:three}). At marginal
stability, hydrodynamical stresses vanish (the apparent singularities
of $T^*_{R\phi}$ in Fig.~\ref{fig:three} do not apply to the
dimensional stress, $T_{R\phi}$, as discussed in \S3.1). The dotted
line in figure~\ref{fig:three} shows the linear perturbation stress
for a representative "hydro-like" mode in a magnetized disk, i.e. a
mode with $(\kva/\Omega)^2=10^{-2}$ which is subject only to weak
magnetic tension.  Even for this "hydro-like" mode, whose properties
become very similar to those of purely hydrodynamical modes as $R
\rightarrow 2 R_g$, the linear perturbation stress remains strictly
positive and continuous throughout the region of interest (as does the
growth rate; see Fig.~\ref{fig:one}).

\subsection{Waves in Magnetized Disks}

While it is not the main focus of this paper, it is worth emphasizing
here what can be learned from the stable branch of the dispersion
relation (eq.~[\ref{eq:ds}]). Indeed, following early investigations
by Kato \& Fukue (1980) and Okasaki, Kato \& Fukue (1987), it has been
suggested that global hydrodynamical modes in unmagnetized accretion
disks could be trapped in the vicinity of black holes (Novak \&
Wagoner 1991; 1992; 1993; Perez et al. 1997; Kato 1990; 1993; 2002;
2003; Li, Goodman \& Narayan 2003). Mode trapping in this context
relies on a purely relativistic effect: the epicyclic frequency
reaches a maximum, before vanishing at the classical marginally-stable
orbit, as one goes deeper and deeper into the black hole potential
well (or that of a very compact neutron star for that matter; see,
e.g., Wagoner 1999 for a review). One of main motivations for
elucidating the properties of such diskoseismic modes is that they may
be at the origin of a variety of quasi-periodic oscillations (QPOs)
observed in compact accreting systems (see, e.g., van der Klis 2000
for a review on QPO observations).

Given the restricted geometry considered, as well as the local and
axisymmetric nature of the analysis, our dispersion relation
(eq.~[\ref{eq:ds}]) describes only a very small subset of all
diskoseismic modes discussed in the literature and global trapped
modes cannot be studied explicitly. Nonetheless, the dispersion
relation is interesting because it does describe, in the magnetized
context, inertio-gravity waves\footnote{We adopt the standard
nomenclature of geophysical fluid dynamics by referring to (internal)
gravity waves in a rotating medium as inertio-gravity waves (see,
e.g., Holton 1992).} which are often regarded as the most
observationally relevant diskoseismic mode (Wagoner 1999).

\begin{figure}
\plottwo{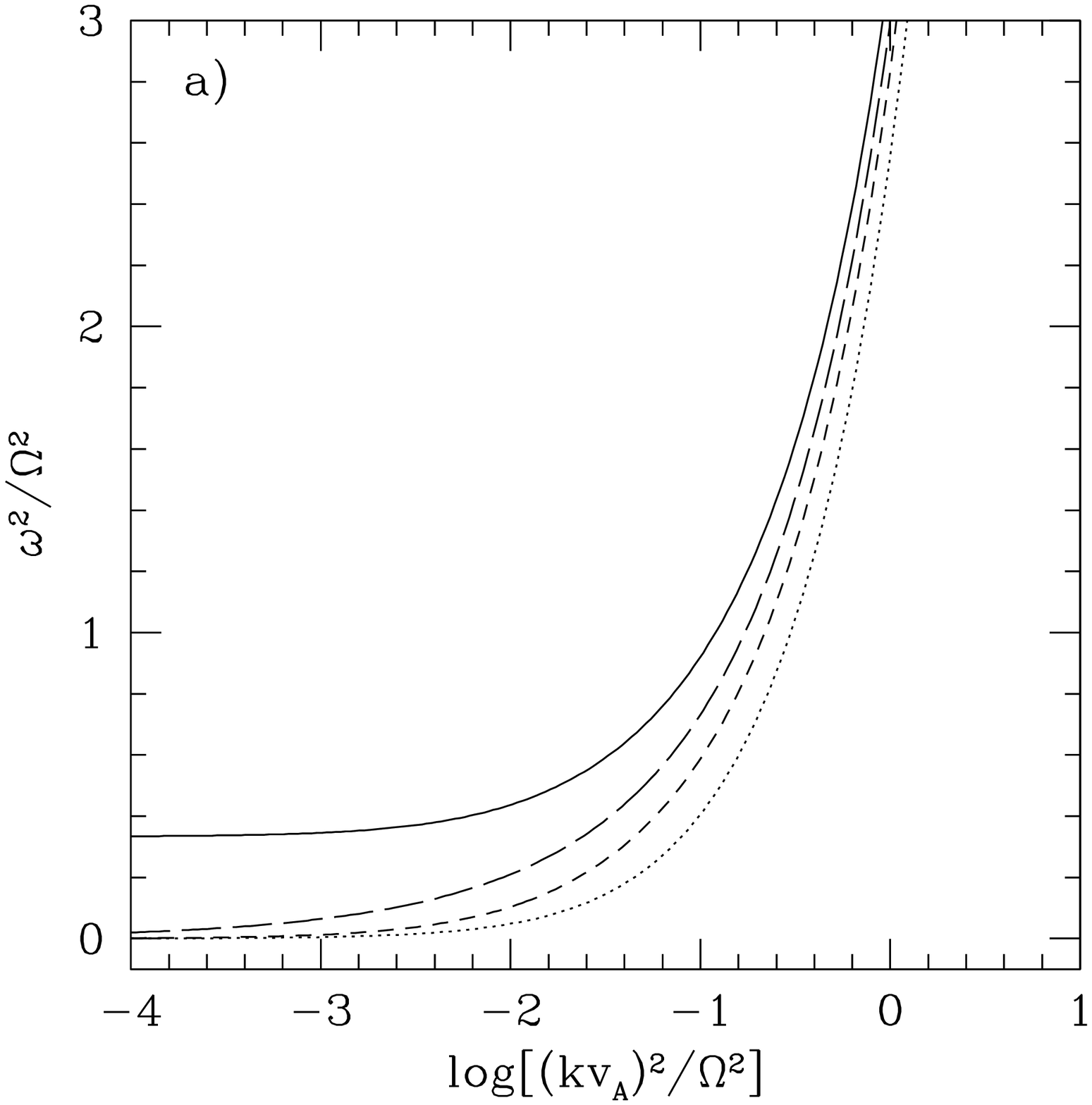}{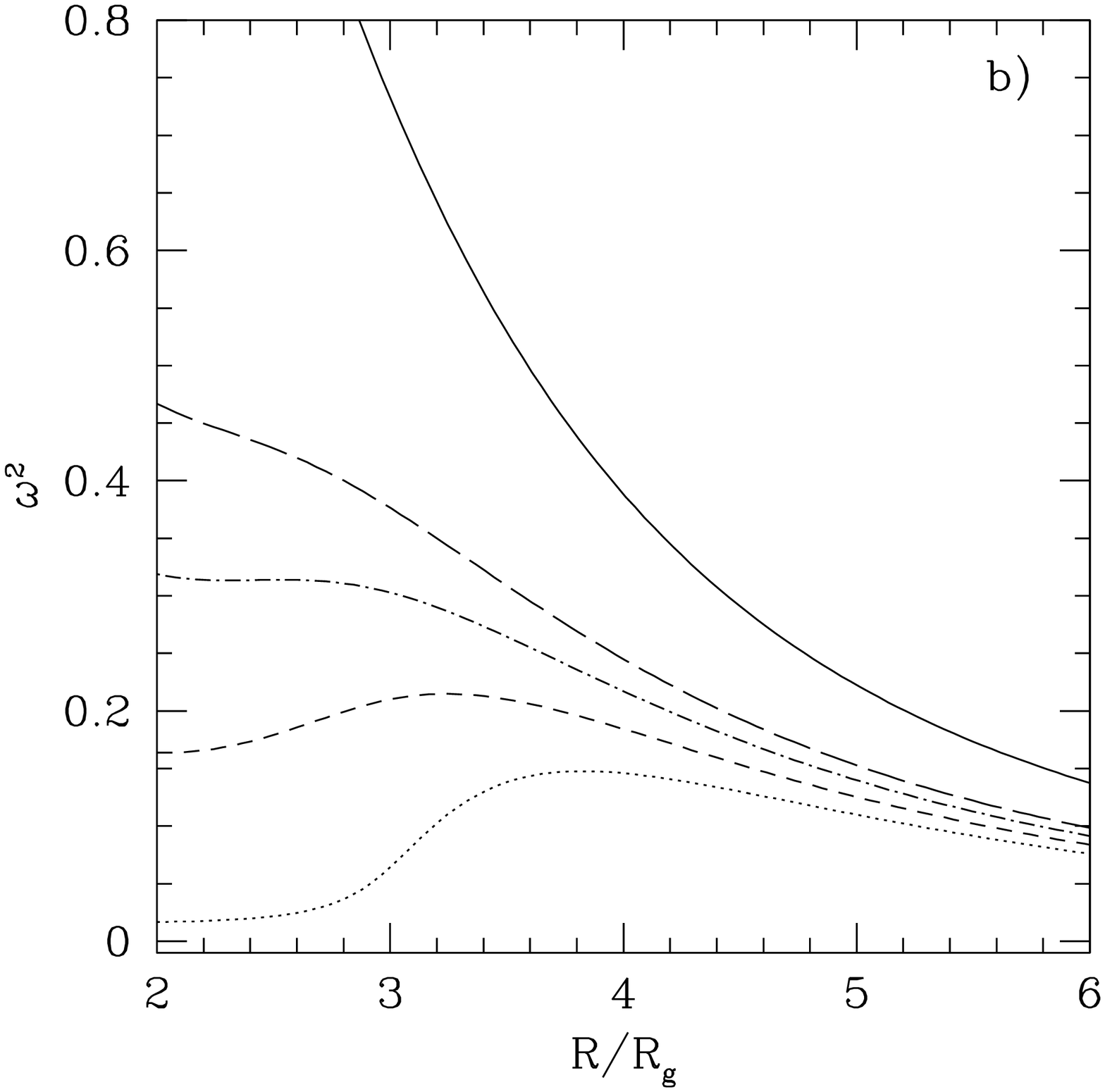}
\caption{(a) Dimensionless frequency, $\omega / \Omega$, of
inertio-gravity-Alfv\`en waves as a function of dimensionless
wavenumber, $\kva / \Omega$, in a magnetized disk without radial
stratification. The various curves correspond to different locations
in the vicinity of the classical marginally-stable orbit (same
notation as Fig.~\ref{fig:one}). Long-wavelength inertio-gravity-like
waves have nearly-zero (but strictly positive) frequencies inside of
the classical marginally-stable orbit. The behavior of
short-wavelength Alfv\`en-like waves, on the other hand, is largely
unaffected by location. (b) Dimensional frequency, $\omega$, of
inertio-gravity-Alfv\`en waves with specific wavenumbers as a function
of location, $R$, in the vicinity of the classical marginally-stable
orbit ($R_{\rm ms}=3 R_g$). From top to bottom, the various curves
correspond to $(\kva)^2/\Omega^2=10^{-1}$, $3 \times 10^{-2}$, $2
\times 10^{-2}$, $10^{-2}$ and $10^{-3}$. The orbital frequency has
been normalized to $\Omega^2=1$ at $R=3 R_g$ in this plot and there is
no radial stratification. Notice that, contrary to the purely
hydrodynamical case, the position of the frequency maximum is not
fixed at $R=4 R_g$ but it varies with wavenumber; the frequency
maximum disappears as magnetic tension becomes important.
\label{fig:four}}
\end{figure}

Figure~\ref{fig:four}a shows the frequency, $\omega$, of wave
solutions to our dispersion relation, as a function of the
dimensionless wavenumber, $\kva / \Omega$. Hydrodynamical analogues to
these waves can be identified by taking the limit $\kva \rightarrow
0$. Outside of the hydrodynamical radius of marginal stability, we
find that the asymptotic value of the wave frequency is $\omega
\rightarrow \sqrt{\kappa^2 + N^2}$ when $\kva \rightarrow 0$.
Clearly, these waves correspond to the (incompressible)
inertio-gravity waves of hydrodynamical theory.\footnote{Note that
because of the restricted geometry adopted (with only $k_Z$ finite),
the analogues of hydrodynamical diskoseismic waves with $\omega <
\kappa$ (absent any radial stratification) are not described by our
dispersion relation.}  As expected, the effect of introducing a finite
radial stratification (not shown here) is to increase or decrease the
wave frequency relative to the epicyclic frequency, $\kappa^2$ (in
addition to changing the location of marginal stability).  However, it
is obvious from figure~\ref{fig:four} that the behavior of these waves
is very different in the large-$\kva$ limit, i.e. when magnetic
tension becomes important. Since we are working in the Boussinesq
approximation, fast magnetosonic waves are effectively filtered out
from the dispersion relation. We also know that unstable modes in
differentially-rotating disks are related to the slow MHD mode, so the
waves described by figure~\ref{fig:four} must correspond to the last
of the three classes of MHD waves, namely Alfv\`en waves (as expected
from the larger frequencies of these waves relative to the frequencies
of slow MHD waves -- described by the other branch of our dispersion
relation -- in the limit $\kva / \Omega \gg 1$; see Balbus \& Hawley
1998 for a discussion of MHD waves in the accretion disk context).

Figure~\ref{fig:four}b illustrates two interesting properties of
inertio-gravity-Alfv\`en waves\footnote{Note the following interesting
property of inertio-gravity modes in a magnetized medium. When they
are stable, these modes are related to Alfv\`en waves, as we have just
discussed. However, when they are unstable, these modes switch branch
in our dispersion relation and become then related to slow MHD
modes. This can be seen by comparing figures~\ref{fig:one}
and~\ref{fig:four} or, equivalently, by comparing the two (stable and
unstable) branches of the dispersion relation in the limit $\kva
\rightarrow 0$, both inside and outside of the hydrodynamical radius
of marginal stability.} in the vicinity of the hydrodynamical radius
of marginal stability. While inertio-gravity waves become purely
evanescent within $R_{\rm ms}$ in unmagnetized disks, the
corresponding waves in a magnetized disk continue to exist within
$R_{\rm ms}$, albeit with very low frequencies (they become evanescent
only asymptotically, i.e. in the limit $\kva \rightarrow 0$). Even if
these waves do not become strictly evanescent ($\omega \rightarrow 0$)
inside of $R_{\rm ms}$, the long-wavelength behavior they exhibit is
broadly consistent with the hydrodynamical limit in the sense that
there is a maximum and a decrease of the wave frequency as one
approaches $R_{\rm ms}$. Note that, contrary to the purely
hydrodynamical case, the location of the frequency maximum is not
fixed at $R=4 R_g$ in magnetized disks but it becomes a function of
wavenumber. Waves with different wavenumbers would then be trapped in
different regions of the disk.

All this suggests that some of the results on trapped inertio-gravity
waves in hydrodynamical disks (due to the presence of a maximum of the
epicyclic frequency, $\kappa^2$) may remain valid in magnetized disks,
provided one is allowed to consider long enough wavelengths for $\kva
/ \Omega$ to be $\ll 1$ (a condition which depends on the magnetic
field strength explicitly). On the other hand, when magnetic tension
becomes important (i.e. if $\kva / \Omega$ is not small),
inertio-gravity-Alfv\`en waves behave much more like Alfv\`en
waves. Figure~\ref{fig:four}b shows that, in that regime, there is no
significant change in the wave properties as one crosses the
hydrodynamical radius of marginal stability. The frequency maximum
disappears and, as a result, mode trapping is no longer expected.

Our short digression on inertio-gravity-Alfv\`en waves indicates that
a more general treatment of waves in magnetized disks is much needed
to address the issue of global mode trapping (including
compressibility effects allowing the existence of fast magnetosonic
waves). Our simple analysis suggests that results established in the
hydrodynamical limit do not carry over automatically to magnetized
disks, especially when magnetic tension becomes important. Another
difficulty that becomes apparent when studying magnetized disks is
that all the waves coexist with the unstable slow MHD
(magneto-rotational) mode, so that ignoring the resulting turbulence
in the treatment of the background disk may be a serious
oversimplification.

\section{Discussion and Conclusion}

The classical description of thin accretion disks around black holes
involves the action of an "anomalous viscosity" far from the hole,
which (slowly) brings gas down to the marginally stable orbit, $R_{\rm
ms}$. At that point, particle orbits become unstable and the gas is
expected to flow into the hole on essentially free-fall orbits. This
rapid infall, whose origin is the dynamical instability at $R_{\rm
ms}$, must then cause the (steady-state) gas density within $R_{\rm
ms}$ to be very small, so that fluid stresses should essentially
vanish. Another reason why one expects stresses to disappear somewhere
outside of the black hole event horizon is that the infall speed must
reach the speed of light at the horizon according to general
relativity and therefore the flow must lose dynamical contact with
external regions once it crosses the sonic (or fast-magnetosonic)
surface.

Several elements of the classical picture should be revised now that
we have a deeper understanding of the process responsible for
accretion in the disk (i.e. the "anomalous viscosity" of the classical
theory). Indeed, the magneto-rotational instability is a dynamical
instability itself. It relies on a rotational imbalance that is also
what fundamentally leads to free-fall within the marginally stable
orbit according to classical theory. Clearly, the fact that magnetized
Keplerian accretion disks are subject to the same dynamical,
magneto-rotational instability far away from the black hole and within
the classical marginally stable orbit, and the realization that this
instability is responsible for accretion far from the hole in the
first place, suggest that the classical marginal stability theory may
be inadequate to describe magnetized accretion disks.

The linear calculations presented in this paper are an attempt to
formulate these objections in a more rigorous way. There is no doubt
that the classical stability theory is exact for test particles (or
dust disks). Hydrodynamical calculations show, however, that positive
linear perturbation stresses resulting from rotational imbalance
(Rayleigh instability) in a nearly-Keplerian gaseous accretion disk
are possible within (and only within) the hydrodynamical radius of
marginal stability. This result is at odds with the classical
description which postulates that the ``anomalous viscosity'' operates
in the disk outside of $R_{\rm ms}$ but not inside of $R_{\rm ms}$.
Even the hydrodynamical analysis may have little to do with
astrophysical systems, however, since those are expected to contain
magnetized disks. Once a magnetic field is introduced, black hole
accretion disks become dynamically unstable to the same
magneto-rotational modes everywhere. Disks are more unstable within
the hydrodynamical radius of marginal stability because of the
increasing shear and ever more destabilizing epicyclic frequency, but
there appears to be no reason (at least in linear theory) why one
should invoke a sharp transition at or about the classical marginally
stable orbit, as illustrated in figure~\ref{fig:three}.

It is interesting that this conclusion is independent of the disk
geometrical thickness.  Indeed, the only requirement for our analysis
to be valid (within ideal MHD) is that the magnetic field be weak
enough for the wavelength of at least one unstable mode to fit within
the disk scaleheigth.  This is roughly equivalent to the requirement
of a subthermal magnetic field ($v_A < c_S$; see, e.g., Balbus \&
Hawley 1998). In our analysis, one is therefore free to chose the disk
sound speed (or equivalently disk thickness) as small as desired, as
long as it remains finite. The results remain unchanged, provided the
basic state magnetic field in the disk is made weak enough to allow
for the presence of unstable modes.

There have been arguments presented in the literature for and against
the classical picture, and in particular related to the issue of
vanishing stresses within $R_{\rm ms}$ (zero-torque boundary
condition). Krolik (1999; see also Agol \& Krolik 2000) and Gammie
(1999) have argued that stresses could remain large across $R_{\rm
ms}$ in a magnetized disk, while Paczynski (2000) and Afshordi \&
Paczynski (2003) have argued in favor of the (near-)vanishing of
stresses at radii $R < R_{\rm ms}$, provided the disk is geometrically
thin (i.e. $H/R \ll 1$). Qualitatively, our linear analysis indicates
that stresses should not vanish exactly anywhere around $R_{\rm ms}$
in a thin, magnetized disk, and in that sense it supports the claims
of Krolik (1999) and Gammie (1999). As a corollary, it suggests that
the classical $R_{\rm ms}$ location loses much of its significance in
magnetized disks.  However, to the extent that the zero-torque issue
is a quantitative one (i.e. large vs. small stresses inside of $R_{\rm
ms}$), it is unclear whether linear stability theory can provide a
useful answer. One should also be aware of important differences
between our approach and the methods of previous authors. While these
authors focused on steady-state models (and some of their conclusions
rely on that property), our stability analysis is intrinsically
time-dependent (a characteristic which may be viewed as a plus). On
the other hand, we do not solve the global near-black-hole accretion
problem, and in doing so, we neglect the important role of large
infall velocities near the black hole event horizon ($v_R \rightarrow
c$) which determine the location where dynamical contact is lost (at
the ``sonic'' surface) and where stresses should vanish.

Another interesting property of our linear analysis is that it is
local. The role of a large-scale magnetic field has at times been
invoked to couple regions of the flow within $R_{\rm ms}$ to regions
more distant from the black hole and thus justify the idea that
stresses may not disappear inside of $R_{\rm ms}$ in magnetized disks
(see, e.g., Krolik 1999; Agol \& Krolik 2000).  Our analysis, because
it is local in nature, shows that such a large-scale
"radially-connecting" magnetic field is not required for stresses to
remain finite across the hydrodynamical radius of marginal stability.

Clearly, there are a number of elements of the near-black-hole
accretion problem that the linear analysis cannot capture. Most
importantly, it is the {global} MHD turbulence resulting from the
non-linear development of the magneto-rotational instability that is
ultimately relevant for the observational properties of accreting
black holes (see Krolik \& Hawley 2002 for a discussion of the issues
in that regime). It is also significant that, while magnetized
turbulent accretion disks have been shown to remain relatively close
to rotational equilibrium (i.e. Keplerian) at large distances from the
black hole (see, e.g., Hawley, Balbus \& Stone 2001), numerical
simulations show (and the infall constraint $v_R \rightarrow c$
guarantees) that accretion must be far from rotational equilibrium in
the vicinity of the black hole event horizon (see, e.g., Hawley \&
Krolik 2001; 2002). This is an indication that linear perturbation
results based on a nearly-Keplerian basic state must be interpreted
with caution in that region. One should also keep in mind that our
analysis was carried out for strictly axisymmetric perturbations and
that non-axisymmetric perturbations may, in general, have different
stability and transport properties. Nevertheless, our results suggest
that there is no reason for stresses to vanish in the vicinity of the
classical marginally-stable orbit, irrespective of the geometrical
thickness of thin accretion disks and whether large-scale
"radially-connecting" magnetic fields are present or not.

\section*{Acknowledgments}

It is a pleasure to thank Steve Balbus and John Hawley for useful
discussions during the completion of this work and Jeremy Goodman for
helpful comments on the manuscript. The author acknowledges support
from the Celerity Foundation.

\clearpage

\clearpage

\clearpage

\begin{figure}
\plotone{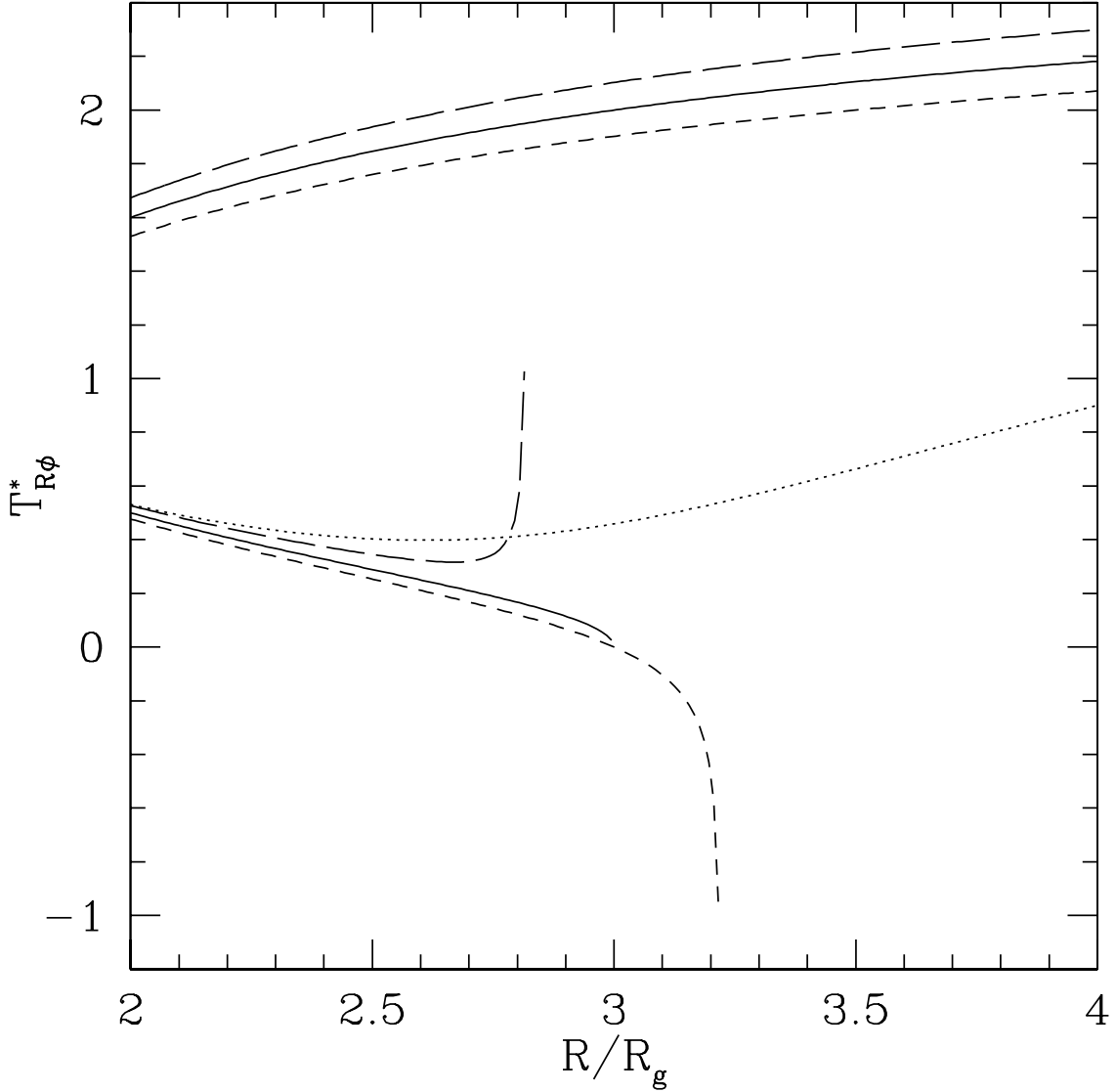}
\caption{Dimensionless stress, $T^*_{R\phi}$, of linearly unstable
modes as a function of location, $R$, in the vicinity of the classical
marginally-stable orbit ($R_{\rm ms}=3 R_g$), for both magnetized and
unmagnetized disks. The two groups of three curves represent stresses
(i) for the fastest growing mode in a magnetized disk (top 3 curves)
and (ii) for (wavenumber independent) unstable modes in a
hydrodynamical disk (bottom 3 curves, interrupted at or around $3
R_g$). In each group, the solid line is for the case without radial
stratification ($N^2=0$), the long-dashed line is for a stabilizing
radial stratification ($N^2=0.1 \Omega^2$) and the short-dashed line
is for a destabilizing radial stratification ($N^2=-0.1
\Omega^2$). The dotted line, which represents the stress for a
long-wavelength mode with $(\kva/\Omega)^2=10^{-2}$ in a magnetized
disk, remains continuously positive throughout the region of interest
even for this "hydro-like" mode. Note that, while signs are accurate,
stress singularities for the bottom short- and long-dashed lines are
only apparent: one can show that the {\it dimensional} stress
vanishes, rather than diverges, at marginal stability (see text for
details).
\label{fig:three}}
\end{figure}

\clearpage

\begin{references}

\reference{} Afshordi, N. \& Paczynski, B. 2003, astro-ph/0202409

\reference{} Agol, E. \& Krolik, J. H. 2000, ApJ, 528, 161

\reference{} Araya-Gochez, R. A. 2002, MNRAS, 337, 795 [Erratum: 2003,
MNRAS, 339, 912]

\reference{} Armitage, P. J., Reynolds, C. S. \& Chiang, J. 2001, ApJ, 548, 868

\reference{} Balbus, S.~A. 2003, ARA\&A, in press (preprint) 

\reference{} Balbus, S.~A. \& Hawley, J.~F. 1991, ApJ, 376, 214 


\reference{} Balbus, S.~A., \& Hawley, J.~F. 1998, Rev.~Mod.~Phys., 70, 1


\reference{} Balbus, S.~A. \& Hawley, J.~F. 2002, ApJ, 573, 749

\reference{} Blaes, O. M. 2003, in "Accretion Disks, Jets, and High
Energy Phenomena in Astrophysics", Proceedings of Session LXXVIII of
Les Houches Summer School, F. Menard, G. Pelletier, G.Henri,
V. Beskin, and J. Dalibard eds. (EDP Science: Paris and Springer:
Berlin)

\reference{} Branduardi-Raymont, G. et al. 2001, A\&A, 365, L140

\reference{} Dabrowski, Y., Fabian, A. C., Iwasawa, K., Lasenby,
A. N. \& Reynolds, C. S. 1997, MNRAS, 288, L11

\reference{} Gammie, C. F. 1999, ApJ, 522, L57

\reference{} Gammie, C. F. \& Popham, R. 1998, ApJ, 498, 313

\reference{} Hawley, J. F., Balbus, S.~A. \& Stone, J. M. 2001, ApJL,
554, L49

\reference{} Hawley, J. F. \& Krolik, J. H. 2001, ApJ, 548, 348

\reference{} Hawley, J. F. \& Krolik, J. H. 2002, ApJ, 566, 164

\reference{} Holton, J. R. 1992, ``Introduction to Dynamic
Meteorology'' (Academic Press: San Diego)

\reference{} Iwasawa, K. et al. 1996, MNRAS, 282, 1038 

\reference{} Kato, S. 1990, PASJ, 42, 99

\reference{} Kato, S. 1993, PASJ, 54, 219

\reference{} Kato, S. 2002, PASJ, 54, 39
 
\reference{} Kato, S. 2003, PASJ, 55, 257

\reference{} Kato, S. \& Fukue, J. 1980, PASJ, 32, 377

\reference{} Krolik, J. H. 1999, ApJ, 515, L73

\reference{} Krolik, J. H. \& Hawley, J. F. 2002, ApJ, 573, 754 

\reference{} Li, L.-X., Goodman, J. \& Narayan, R. 2003, preprint
(astro-ph/0210455)

\reference{} Miller, J. M. et al. 2002, ApJ, 570, L69

\reference{} Narayan, R., Quataert, E., Igumenshchev, I.~V. \&
Abramowicz, M.~A., 2002, ApJ, 295, 301

\reference{} Nowak, M. A. \& Wagoner, R. V. 1991, ApJ, 378, 656
 
\reference{} Nowak, M. A. \& Wagoner, R. V. 1992, ApJ, 393, 697

\reference{} Nowak, M. A. \& Wagoner, R. V. 1993, ApJ, 418, 187 

\reference{} Novikov, I. D.  \& Thorne, K. S. 1973, in Black Holes,
ed. C. de Witt \& B. de Witt (New York: Gordon \& Breach), p. 343

\reference{} Okazaki, A. T., Kato, S. \& Fukue, J. 1987, PASJ, 39, 457

\reference{} Paczynski, B. 2000, astro-ph/0004129

\reference{} Paczynsky, B. \& Wiita, P. J. 1980, A\&A, 88, 23

\reference{} Page, D.N. \& Thorne, K. S. 1974, ApJ, 191, 499

\reference{} Perez, C. A., Silbergleit, A. S., Wagoner, R. V. \& Lehr,
D. E. 1997, ApJ, 476, 589

\reference{} Reynolds, C. S. \& Armitage, P. J. 2001, ApJ, 561, L81 

\reference{} Reynolds, C. S. \& Begelman, M.C. 1997, ApJ, 488, 109

\reference{} Shapiro, S. L. \& Teukolsky, S. A. 1983, Black Holes,
White Dwarfs and Neutron Stars -- The Physics of Compact Objects (New
York: Wiley-Interscience)

\reference{} van der Klis, M. 2000, ARA\&A, 38, 717

\reference{} Wagoner, R. V. 1999, Phys. Rep., 311, 259  

\reference{} Wilms, J. et al. 2001, MNRAS, 328, L27

\reference{} Young, A. J., Ross, R. R. \& Fabian, A. C. 1998, MNRAS,
300, L11

\end{references}
\end{document}